\documentclass[journal=accacs,manuscript=article,layout=onecolumn]{achemso}
\usepackage[version=3]{mhchem} 

\usepackage{macros}
\usepackage{header}
\glsaddall

\abbreviations{}

\let\oldmaketitle\maketitle
\let\maketitle\relax

\begin{document}



\oldmaketitle

\begin{abstract}
Access to the potential energy Hessian enables determination of the Gibbs free energy, and certain approaches to transition state search and optimization. Here, we demonstrate that off-the-shelf pretrained Open Catalyst Project (OCP) machine learned potentials (MLPs) determine the Hessian with great success (58 cm$^{-1}$ mean absolute error (MAE)) for intermediates adsorbed to heterogeneous catalyst surfaces. This enables the use of OCP models for the aforementioned applications. The top performing model, with a simple offset correction, gives good estimations of the vibrational entropy contribution to the Gibbs free energy with an MAE of 0.042 eV at 300 K. The ability to leverage models to capture the translational entropy was also explored. It was determined that 94\% of randomly sampled systems had a translational entropy greater than 0.1 eV at 300 K. This underscores the need to go beyond the harmonic approximation to consider the entropy introduced by adsorbate translation, which increases with temperature. Lastly, we used MLP determined Hessian information for transition state search and found we were able to reduce the number of unconverged systems by 65\% to 93\% overall convergence, improving on the baseline established by CatTSunami.

\end{abstract}

\section{Introduction}




The potential energy Hessian captures the curvature of the \gls{PES}. The Hessian may be diagonalized to find the vibrational modes and frequencies of atomic species. Knowledge of the vibrational frequencies allows the vibrational contributions of the entropy to be estimated as well as the \gls{zpe}. Knowledge of the \gls{zpe} and the entropy are necessary to calculate the free energy. When assessing whether a reaction is thermodynamically favorable, the free energy of \textit{reaction} is considered. For constant pressure systems more particularly, the change in the Gibbs free energy is considered. To determine the rate of reaction, knowledge of the free energy of \textit{activation} is necessary. In this sense, it is imperative to calculate the potential energy Hessian when seeking to answer the core questions of catalysis: the thermodynamic favorabilities and rates of reactions. 

In heterogeneous catalysis, it is common to use the harmonic approximation, which assumes that the potential is harmonic about the minimum energy points. It is common to also assume the entropy is well treated by only considering the vibrational contribution to it, and that the vibrational energy from the surface may be neglected\cite{vojvodic2014exploring, amsler2021anharmonic, jorgensen2018monte, norskov2005trends, HTS_w_ML_2017}. These approximations neglect the translational, rotational, and electronic contributions to the entropy. The translational entropy is particularly important if the adsorbate is able to freely move about the surface. The harmonic approximation is poor at high temperatures where access to anharmonic energized states is possible and for more complex systems like those with solvent, zeolites, porous materials, and supported nanoparticles\cite{collinge2020Effect, Piccini2022Ab} where other contributions also become important.
There have been several proposed approaches to supplement the harmonic approximation. The hindered translator (and hindered rotor) model\cite{sprowl2016hindered} was introduced to account for the translational (and rotational) entropy of adsorbed species, obtained from partition functions associated with 1D surface diffusion and rotations hindered by activation energies. The \gls{cpes} method\cite{jorgensen2017Adsorbate} models instead the translational entropy from the explicit calculation of the 2D potential energy experienced by the molecules adsorbed to the surfaces, calculated with a series of constrained relaxations. Adsorbate entropies can also be estimated from the distribution of states extracted from molecular dynamics (MD) trajectories. In the quasi-harmonic approximation method\cite{piccini2014effect}, anharmonicity is accounted for by calculating the vibrational density of states using the velocities extracted from MD simulations. Anharmonic corrections can be computed also with the thermodynamic integration method, from the harmonic to the fully interacting system\cite{amsler2021anharmonic}. Alternatively, adsorbate entropies can be obtained from enhanced sampling methods such as Metadynamics or the Blue-Moon approach\cite{Piccini2022Ab, barducci2011Metadynamics, ciccotti2004blue}. These approaches also enable the calculation of the entropy of transition states from MD\cite{debnath2019enhanced}.

Since the release of \gls{oc20}\cite{chanussot2021open}, there has been momentum in training novel \gls{ML} model architectures\cite{schütt2017schnetcontinuousfilterconvolutionalneural, schütt2021equivariantmessagepassingprediction, gasteiger2020fast, gasteiger2022gemnet, liao2023equiformerv2} to obtain versatile \gls{mlp}\cite{clausen2024adapting} that may be used to accelerate or even replace \gls{dft} in computational investigation of heterogeneous catalyst systems\cite{lan2023adsorbml, wander2024cattsunami}.

In some cases, pretrained \gls{mlp}s may benefit from additional fine-tuning on specific tasks. 
Fine-tuning is a training technique applied to pretrained models, which is commonly used in fields such as natural language processing to adapt neural network models to tasks outside the domain of the original pretraining data set \cite{devlin2019bertpretrainingdeepbidirectional}.
For molecular property prediction, fine-tuning has been shown to be effective both for improving the accuracy of previously learned properties for systems which are outside of the domain of the training set, and for learning to predict new properties while benefiting from useful latent information learned from the previous domain \cite{zaidi2022pretrainingdenoisingmolecularproperty, shoghi2024moleculesmaterialspretraininglarge, Wang_2024}. 
Predicting the numerical Hessian of the potential energy as the Jacobian of the forces relies on \gls{mlp} predictions of the forces, which pretrained \gls{oc20} \gls{mlp}s were trained to predict. The potential energy Hessian is approximated numerically using finite differences.
The displacements of atoms used in finite differences are at a length scale which are not necessarily represented in the \gls{oc20} dataset, because the dataset primarily contains relaxation trajectories of adsorbate catalyst systems \cite{chanussot2021open}.
Therefore it may be valuable to fine-tune pretrained \gls{mlp}s on samples of the small displacements used to calculate Hessians numerically, so that the effect of these small displacements are adequately represented in the neural network.

\begin{figure}
    \centering
    \includegraphics[width=0.87\linewidth]{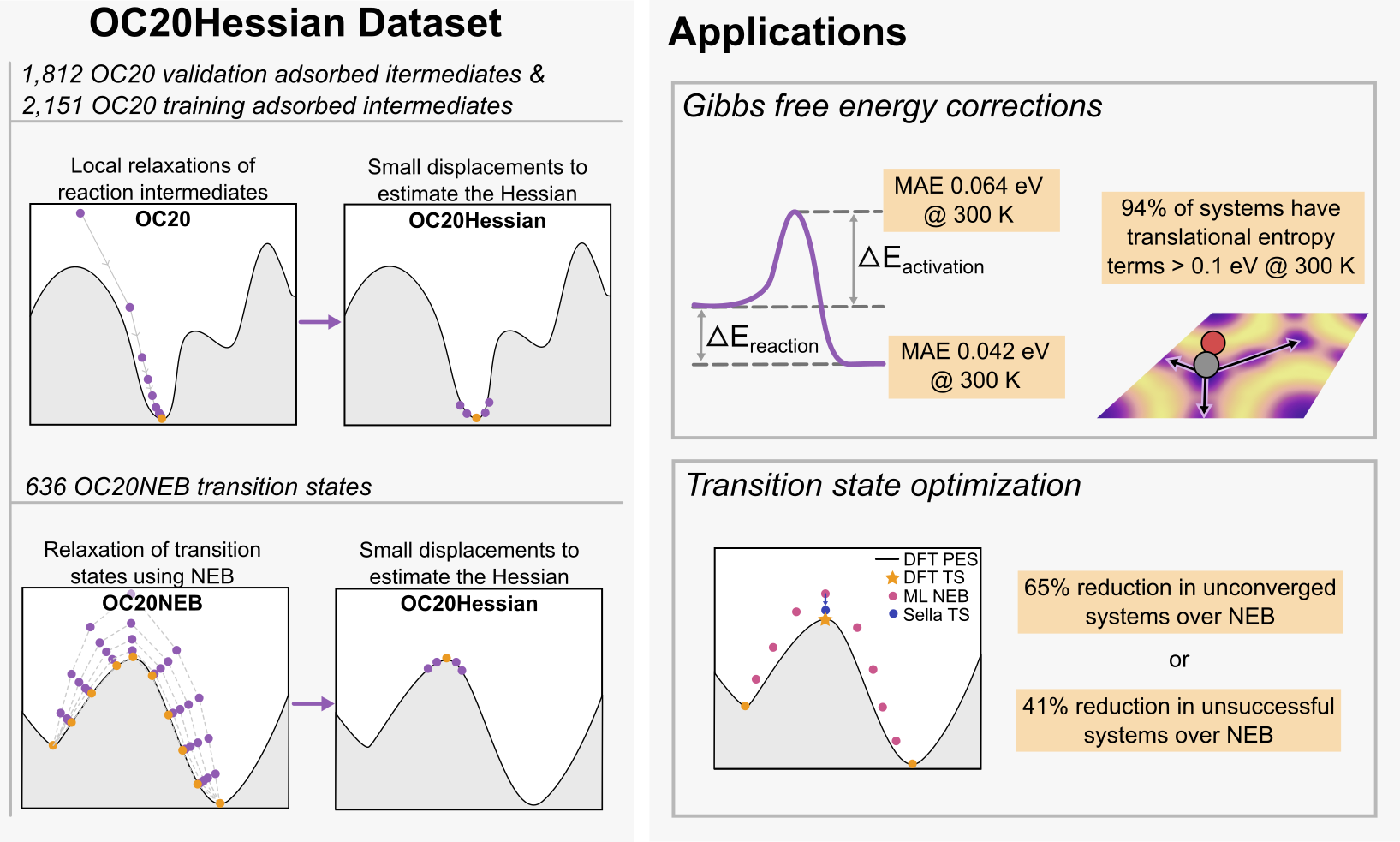}
    \caption{A summary of the work presented. We demonstrate that pre-trained graph neural networks are able to determine the potential energy Hessian with high precision. This allows the Gibbs free energy corrections to be calculated with low errors. It is determined that many systems have significant translational entropy contributions at 300 K. The Hessians are also used for transition state optimization with Sella which greatly improves convergence.}
    \label{fig:summary}
\end{figure}

Here, we seek to answer four outstanding questions about the use of \gls{oc20} trained \gls{mlp}s, which are important to consider for practical use of the models in computational studies: (1) Can \gls{oc20} trained models be used to obtain vibrational frequencies accurately? (2) Can \gls{oc20} trained models be used to inexpensively augment entropy approximations? (3) How should Gibbs free energies most easily be accessed? (4) Can \gls{oc20} trained models be used to obtain energy Hessians that are useful in transition state optimization? The important components and outcomes of this inquiry are summarized in Figure \ref{fig:summary}.

To address these questions, we constructed a dataset of 4,599 numerical Hessians using \gls{dft}: 1,812 randomly sampled relaxed structures from the \gls{oc20} validation dataset, 2,151 randomly sampled relaxed structures from \gls{oc20} training dataset, and 636 transition states from the \gls{OC20NEB}\cite{wander2024cattsunami}. Using the calculated numerical Hessians, we assessed the performance of five pretrained models for determination of the numerical Hessian and found good agreement between \gls{dft} vibrational modes and those determined using the \gls{mlp}. A \gls{mae} of 58 cm$^{-1}$ was achieved by the top performing model on adsorbed intermediates. Using the \gls{ML} and \gls{dft} vibrational modes, the Gibbs free energy corrections at 300K were considered to ground the impact of \gls{ML} errors on the resultant Gibbs free energy. For the top performing model, the \gls{mae} was determined to be 0.042 eV. This outperforms the commonly used approach of calculating the Gibbs correction for a particular adsorbate on one surface and assuming that value holds true for all surfaces, which has an \gls{mae} of 0.058. The opportunity of fine-tuning was also explored, which reduced the \gls{mae} further to 0.031 eV.

For 100 (of 1,812) systems from the validation dataset, the \gls{cpes}\cite{jorgensen2017Adsorbate} approach was applied using \gls{mlp}s, reducing entropy approximations to include the translational degrees of freedom for the adsorbed species. Across the 100 systems considered, the average difference in energy brought about by neglecting the translational degrees of freedom was 0.16 eV, with a maximum difference of 0.27 eV at 300 K. This underscores the value added by the low cost of the \gls{mlp}s, which makes broader use of approaches which more completely treat the entropy feasible. Other techniques to go beyond the harmonic approximation such as \gls{AIMD}\cite{lee2021Neural, jiang2020high} or enhanced sampling techniques could also be used with \gls{oc20} pretrained models, but consideration of these approaches was beyond the scope of this work.

To address the easiest method for accessing Gibbs free energy corrections, we considered the approach of using an \gls{ML} potential to compute the Hessian. We also considered the simplest approach: constructing a lookup table with the average frequencies across the DFT dataset. Although using the \gls{mlp} performs the best (0.042 eV \gls{mae} at 300 K) for \gls{oc20} validation data, the lookup table is comparable (0.045 eV \gls{mae} at 300K). The lookup approach does not extend to arbitrary adsorbate, which is especially important when considering transition states. For this reason, we recommend using known corrections when possible, and \gls{ML} determined corrections where otherwise necessary. 

Finally, we performed transition state optimization with Sella\cite{hermes2022sella} on the ML-determined transitions states from the OC20NEB\cite{wander2024cattsunami} dataset using the numerical Hessian from the \gls{mlp}s. By performing this optimization, we were able to increase the convergence rate from 80\% to 93\%. This demonstrates the usefulness of pretrained model Hessians on transition state optimization. It is likely the \gls{mlp} could be applied to other Hessian driven transition state search algorithms such as the growing string method\cite{peters2004growing, zimmerman2013reliable, zimmerman2015single} or earlier transition state search methods that rely on the Hessian\cite{cerjan1981finding, nguyen1985finding, taylor1985imposition, baker1986algorithm, wales1989finding, quapp1996gradient}. Success using \gls{gnn} to perform transition state optimization with Sella has been successfully demonstrated for organic molecules with fine-tuning by Yuan et al.\cite{yuan2024deep}

The contributions of this work are five-fold: (1) creation of a dataset of \gls{dft} 4,599 numerical Hessians, (2) assessment of pretrained model performance at computing numerical Hessians, (3) fine-tuning of models to improve their performance for Hessian determination, (4) testing the ability to leverage \gls{mlp}s to augment the harmonic approximation, and (5) Assessment of an OCP pretrained model for transition state optimization using Sella.

\section{Methods}
\subsection{Generating the dataset}
1,812 systems were sampled from the OC20\cite{chanussot2021open} validation dataset. An additional 2,151 were sampled from the OC20 training dataset. For those sampled from the training dataset, systems were limited to those with fewer than 100 atoms to reduce computational cost. Additionally, we considered all barriered reaction transition states presented in the OC20NEB\cite{wander2024cattsunami} dataset. For each of these sampled systems, a finite differences approach without symmetry simplifications was used to determine the numerical Hessian. 
Calculations were performed using \gls{vasp}\cite{kresse1994ab, kresse1996efficiency, kresse1996efficient, kresse1999ultrasoft} with the revised Perdew-Burke-Ernzerhof (RPBE) functional\cite{perdew1996generalized, zhang1998comment}. 
For each direction, two displacements of 0.0075 \r{A} were performed. The electronic convergence criterion was met if the change in energy between iterations was less than \num{1e-6} $eV$. 
For the OC20 dataset\cite{chanussot2021open}, the electronic convergence criterion was less strict (\num{1e-4} $eV$). 
The finite difference \gls{dft} calculations were also run with an electronic convergence criterion of \num{1e-4} $eV$ for comparison with the \num{1e-6} $eV$ criterion.
Only minute differences were observed between these convergence criteria across all OC20 validation and training systems, despite \num{1e-6} $eV$ being the recommended value. 
Otherwise, all other settings were kept consistent with the OC20 dataset. 
This alleviates any concern that the models would have issues with this task due to this discrepancy.  
For more information about the calculation details and a histogram of the absolute error associated with the different electronic convergence criteria see the Supplementary Information.
The dataset and associated code is available on GitHub at https://github.com/jmusiel/gibby/tree/main.

\subsection{Calculating the Gibbs Energy Corrections}
The Gibbs free energy corrections have three terms which depend on the vibrational modes: the \gls{zpe}, vibrational specific heat term, and the entropy term (TS) as shown in Equation \ref{eqn:gibbs_corr}. The specific heat term (Equation \ref{eqn:cp}) may be neglected because an equal an opposite term appears in the entropy (Equation \ref{eqn:ts}). For this work, we used the HarmonicThermo class in \gls{ase}\cite{larsen2017atomic} to calculate the Gibbs corrections. If an imaginary vibrational mode was found, we discarded those modes in the calculation. These could be zero-valued but are small negative eigenvalues because of the curvature approximation from finite differences. If zero-valued, these should be interpreted as a loss of a vibrational degree of freedom in favor of a translational mode. In this case, to get an accurate entropy, the 2D translational entropy should be calculated. If the actual value is small and real valued, the vibrational mode should be interpreted as a frustrated vibration. In this case the finite differences approach should be altered to better capture the curvature correctly.

For evaluation of the performance of ML models in comparison to explicit calculation we consider two cases (1) mean per adsorbate \gls{dft} and (2) Adopt-a-value \gls{dft}. For mean per adsorbate, the mean ordered frequencies across all systems are computed per adsorbate. Then the error between the Gibbs energy corrections obtained using the actual frequency values and the mean frequency values is computed for every system. For Adopt-a-value, the systems are iteratively considered and exhaustively paired with other systems with the same adsorbate on different surfaces. For each pair, the difference in energy is the error. The error is averaged across all systems. This case is the most similar to the common practice\cite{vojvodic2014exploring,norskov2005trends, HTS_w_ML_2017} of calculating the per adsorbate Gibbs corrections on one surface and taking them to be true across all surfaces.

\begin{equation}\label{eqn:gibbs_corr}
    \Delta G_{corr} = ZPE + \int C_p dT - TS
\end{equation}

\begin{equation}\label{eqn:zpe}
    ZPE =  \sum_{\nu} \frac{1}{2} h \nu
\end{equation}

\begin{equation}\label{eqn:cp}
    \int C_p dT=  \sum_{\nu} \frac{h\nu}{e^{h\nu/kT} - 1}
\end{equation}

\begin{equation}\label{eqn:ts}
    TS =  \sum_{\nu} \frac{h\nu}{e^{h\nu/kT} - 1} - ln(1 - e^{-hv})
\end{equation}

\subsection{Hessian Accuracy Metrics}
The Hessian matrix of the potential energy was predicted for each system in the dataset using a variety of \gls{gnn} \gls{mlp}s with high performance on \gls{oc20} benchmarks.
A numerical finite differences approach was used to obtain the Hessian matrix from the \gls{mlp}, which is the same approach that is used by \gls{vasp}.
The numerical approach was chosen over an automatic analytical approach after testing showed automatic differentiation to be slower than the finite differences approach, with similar or worse accuracy when replicating \gls{vasp} finite difference data.
More information on the choice of finite differentiation instead of analytical differentiation can be found in the Supplementary Information. 

We examined three different metrics for assessing the accuracy of Hessian matrix predictions.
First, we are interested in the overall accuracy of the Hessian prediction itself, as a way of examining the overall efficacy of the \gls{mlp}. The \gls{mlp}s have been trained on force and energy property labels, but here they are used to predict the curvature which is the second derivative of energy, or the first derivative of the forces.
To assess the overall Hessian accuracy, we look at the \gls{mae} of the sorted frequencies.
Alternatively, we could use the sorted eigenvalues, but these scale with the square of the frequencies, and therefore capture the same information.
Second, we are interested in the performance of the Hessian as a tool for predicting the Gibbs free energy correction, which is a function of the frequencies.
Therefore we use the \gls{mae} of \gls{mlp} Hessian predicted Gibbs free energies versus \gls{vasp} Hessian predicted Gibbs free energies.
This metric should generally correspond to the overall accuracy of the Hessian, but may be biased by significant errors in the very high magnitude and very low magnitude frequency ranges.
Third, we are interested in the use of the Hessian to identify transition states versus relaxed states.
Ideally, transition states are characterized by a single imaginary frequency, and therefore the only frequency of interest is the \gls{LI/SR} frequency.
We measured the \gls{mae} of this \gls{LI/SR} frequency to examine the effectiveness of the Hessian prediction for identifying, and optimizing towards, transition states.

\subsection{Fine-tuning Experiments}
To examine the effects of fine-tuning a pretrained Equiformer V2 checkpoint, we used data generated by vibration calculations on local minima from the OC20 training dataset. 
We selected the 153M parameter Equiformer V2 checkpoint for the starting model weights because it is the best performing pretrained checkpoint we tested. 
Instead of training the model to directly predict the Hessian, we fine-tuned it only to predict the same force and energy properties it was already trained for, with a different learning rate on a different data set.

The difference between the fine-tuning step and the pretraining is the data set used for fine-tuning, which contains numerous single point calculations generated by \gls{dft} using the finite differences approach to generate the Hessian.
These single point calculations contain numerous small displacements of the adsorbate atoms, smaller than would typically be found in a relaxation trajectory which composes the original pretraining data set.
To select a learning rate, we performed hyperparameter optimization and determined a learning rate of $5\times 10^{-6}$ to be best.
We validated the model's force and energy performance on a held-out set of relaxed structures to determine when to stop training.
Using this held-out set, we determined an optimal stopping point for this data set after 12 epochs, but we also measured the accuracy on our test sets after each epoch from 0 to 20 to examine the performance trends associated with fine-tuning.

We measured the fine-tuned model's performance in predicting the Hessian and Gibbs correction on a held out test set of Hessians computed for relaxed structures by \gls{dft}, and another test set of Hessians computed for transition states by \gls{dft}.
We measured the Hessian prediction performance by looking at the MAE of the \gls{LI/SR} frequency compared to the corresponding \gls{dft} generated \gls{LI/SR} frequency.
We expect this \gls{LI/SR} frequency to be indicative of the model's ability to accurately optimize transition states versus relaxed states using the diagonalized Hessian.
We also measured the overall Gibbs energy correction error by taking the \gls{mae} of the Gibbs energy correction computed from model predicted Hessian, versus the correction computed from the \gls{dft} Hessian.

Because we fine-tuned the model only on single point calculations from Hessians on relaxed structures, we expect there to be a difference in performance between the Hessian predictions on other relaxed structures (local minima), and Hessian predictions on transition state structures (saddle points).
In particular, we expect to see a significant difference in the performance of the model on the \gls{LI/SR} frequency, which corresponds to the reactive mode.
Therefore we specifically examined the distribution of errors on just the \gls{LI/SR} frequencies, comparing the distribution of the relaxed structures from the OC20 validation set to the saddle points from the OC20NEB transition states set.

\subsection{Complete Potential Energy Sampling}
To facilitate the evaluation of the translational entropy, we created a suite of tools to perform potential energy sampling by the procedure described by J{\o}rgensen et al\cite{jorgensen2017Adsorbate}. 
These tools are available on GitHub at https://github.com/jmusiel/gibby/tree/main.
The tool takes as input the adsorbate, slab, and an \gls{ase}\cite{larsen2017atomic} calculator object to perform constrained relaxations of the adsorbate on the slab. 
There is an \gls{ase} calculator object compatible with evaluating OC20 pretrained models, making their use facile. The constrained relaxations are performed on a grid with adjustable spacing. 
Results are cached so they may be accessed at a later time. 
Once the constrained relaxations are performed, a surrogate \gls{PES} is constructed and used to evaluate the translational partition function and entropy by Equations \ref{eqn:z_translational} and \ref{eqn:S_translational}, which was taken from the work by J{\o}rgensen et al\cite{jorgensen2017Adsorbate}. 
We also created classes similar to the \gls{ase} HarmonicThermo class, to allow for the rapid evaluation of the Gibbs free energy for a constructed \gls{PES} and vibrational modes.

\begin{equation}\label{eqn:z_translational}
    Z_{T}^{CPES} = \frac{2 \pi m k_B T}{h^2} \iint exp \left( \frac{-V(x,y)}{k_B T} \right) dx dy
\end{equation}

\begin{equation}\label{eqn:S_translational}
    S = k_B ln(Z) + k_BT\frac{1}{Z} \left(\frac{\partial Z}{\partial T} \right)_{V,N}
\end{equation}

To validate the \gls{cpes} approach, 100 systems from the 1,812 OC20 validation set were analyzed. For each, system, a mesh grid of points was constructed across the unit cell surface with a spacing of 0.5 \AA{}. 
For each one of these points a constrained relaxation of the adsorbate was performed using the 31M parameter Equiformer v2\cite{liao2023equiformerv2} pretrained model. The adsorbate was prevented from dissociating using Hookean constraints which allowed bond elongation of 1.3$\times$ the relaxed bond length, beyond which a 10 eV/\AA{} force was applied. 
Additionally, all surface atoms were fixed and the adsorbate center of mass was fixed in the directions parallel to the surface, but allowed to move in the direction normal to the surface. 
Using the relaxed energies across the meshgrid, a finer surrogate grid was constructed using cubic spline interpolation. 
The surrogate mesh grid was used to integrate the potential energy across the surface to determine the associated partition function by Equation \ref{eqn:z_translational}. 
With the partition function, the entropy may be readily assessed (Equation \ref{eqn:S_translational}). \gls{dft} single points were also performed on the \gls{ML} relaxed systems. 
These single points were performed with the functional and settings consistent with the OC20 dataset\cite{chanussot2021open}.

\subsection{Transition state optimization with Sella}
Transition state optimization algorithms such as Sella make use of an exact or approximate Hessian matrix to efficiently drive the optimization towards the nearest saddle point \cite{hermes2022sella}.
We measure the effectiveness of Hessian matrix calculations from models trained on \gls{oc20} for performing transition state optimization.
To do this, we examine the effects of using Sella to optimize the transition states identified in the OC20NEB data set.
We observed Sella to be sensitive to hyperparameter initialization, so we performed a hyperparameter optimization to determine the appropriate parameters to be used.
The parameters selected maximized the convergence rate over 500 steps for the 153M parameter Equiformer V2 model on the OC20NEB transition state dataset.
A convergence criterion of $\gamma = 0$ was used to ensure that the Hessian matrix was recalculated and used at every step in the optimization.
A trust radius of $\eta = \num{7e-4}$ $\text{\AA}$ was used.
More details on the specifics of all the hyperarameters used for the Sella optimization can be found in the SI.

Sella's performance as a transition state optimizer is compared to the CatTSunami baseline which identified transition states using the \gls{neb} technique.
For the Sella approach, the ML \gls{neb} optimized transition state structure is refined with Sella to convergence, or for 500 steps, whichever happens first.
Two variants of this approach are compared: (1) unconverged structures from the \gls{neb} approach or from Sella are removed completely to improve accuracy, and (2) where Sella converged structures are prioritized but \gls{neb} converged structures are used in the event that Sella fails to converge to improve convergence rate.
Performance of the transition state optimization is quantified in several ways: maximizing convergence rate, minimizing the \gls{Fmax}, minimizing the error in the \gls{Ea} , and maximizing the success rate.
The convergence rate is the percentage of systems which reach a transition state below \gls{Fmax} criteria (0.01 eV/$\text{\AA}$ for Sella, 0.05 eV/$\text{\AA}$ for \gls{neb}s).
The \gls{Fmax} is the magnitude of the maximum force when a \gls{dft} single point is performed on the \gls{ML}-relaxed transition state.
The \gls{Ea} energy residuals are the differences between the activation energy of the \gls{dft} single point performed on the \gls{ML} optimized transition state, and the result of a \gls{dft} \gls{neb} calculation starting from the same system.
The success rate is the proportion of systems where the activation energy of the \gls{dft} single point performed on the \gls{ML} optimized transition state is accurate to within 0.1 eV of the energy given by the \gls{dft}-optimized transition state from an \gls{neb} calculation.

\section{Results and Discussion}
\subsection{Hessian accuracy from pretrained force models}
\begin{figure}
\centering
\includegraphics[width=0.76\textwidth]{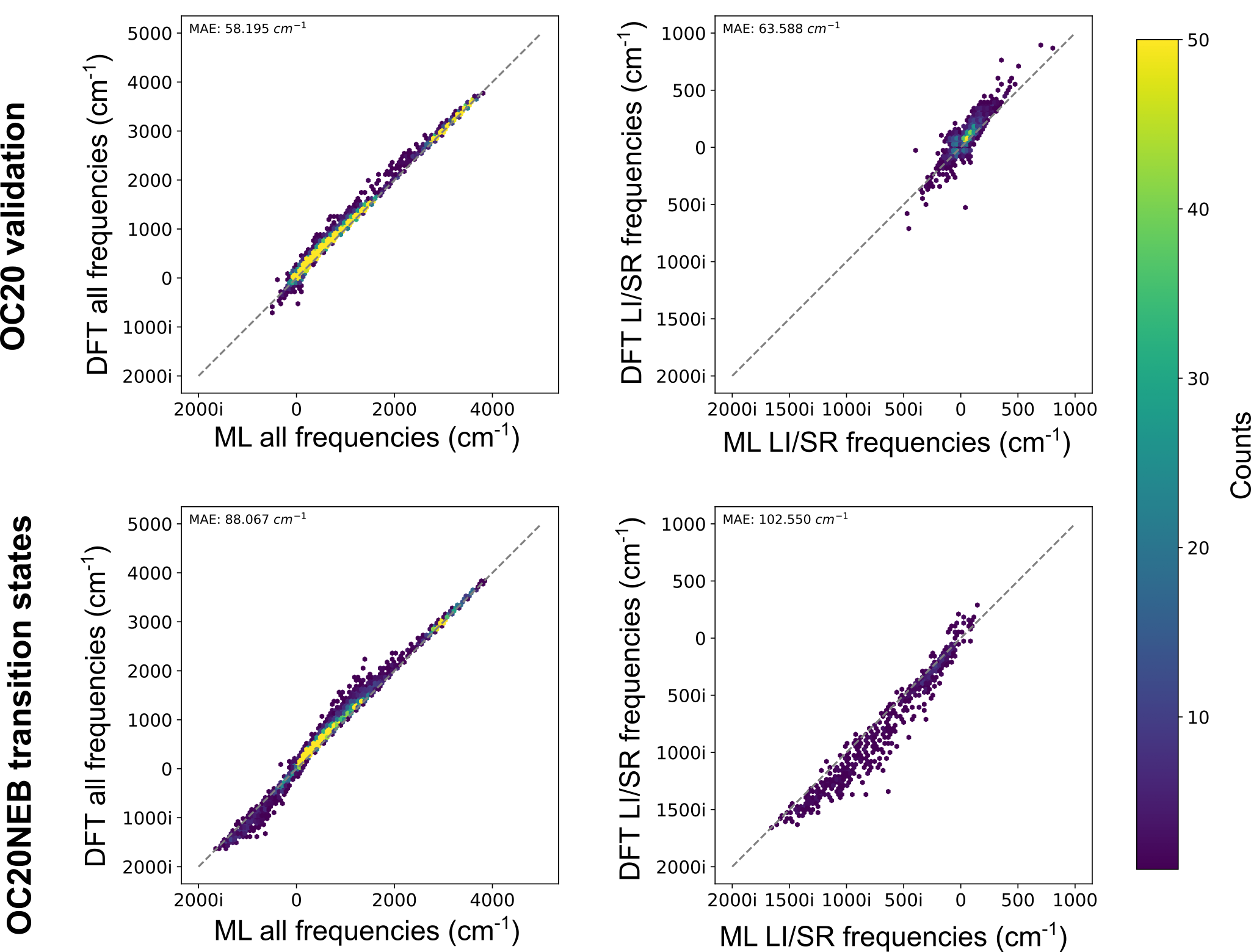}
\caption{
    Parity plots of the frequencies for all vibration modes and the \gls{LI/SR} frequencies of the Hessian predicted by \gls{vasp} versus the Hessian predicted by the 153M parameter Equiformer V2 via finite differences.
}
\label{fig:eq2_hessian_accuracy_parity_plot}
\end{figure}

We measure the accuracy of predicting the Hessian from a pretrained \gls{gnn} by numerically computing the gradients of the forces via finite differences. 
This approach is identical to the one implemented in  \gls{vasp}, which was used to calculate the \gls{dft} numerical Hessians. The performance of the top performing model (Equiformer V2 153M parameters) is shown in Figure \ref{fig:eq2_hessian_accuracy_parity_plot} on the OC20 validation systems (top) and the OC20NEB transition states (bottom). 
For both data types, a parity plot of the vibrational frequencies (left) reveals the high degree of parity between the \gls{dft} and \gls{ML} results. 
There is, however, a small systematic bias.
The models systematically predict larger magnitude eigenvalues and therefore the vibrational frequencies to be slightly larger than the actual values.
Interestingly, this is the opposite effect to that noted by Deng et al.\cite{deng2024overcoming} for three different \gls{mlp}s on out of domain systems.
This bias can and will be corrected for in the application of the Hessian.
Special attention is also paid to the \gls{LI/SR} vibrational frequency (shown on the right). In the case of a transition state, this should correspond to the eigenmode of reaction. 
For the transition state systems, the distribution of the \gls{LI/SR} frequency is shifted to the left when compared to the OC20 validation systems as we would expect. 
For non-reacting systems, there are still some with imaginary eigenmodes. 
In this case, it is likely the curvature is shallow and the imaginary frequency is an artifact of the numerical technique. 
Still, for this important frequency on the OC20NEB transition states, we find there is good parity between the \gls{ML} and \gls{dft} values. 
This is promising for the use of OC20 pretrained \gls{mlp}s for transition state optimizations which rely on the Hessian.

\subsection{Fine-tuning a GNN on vibration singlepoints}

\begin{figure}
    \centering
    \includegraphics[width=0.87\linewidth]{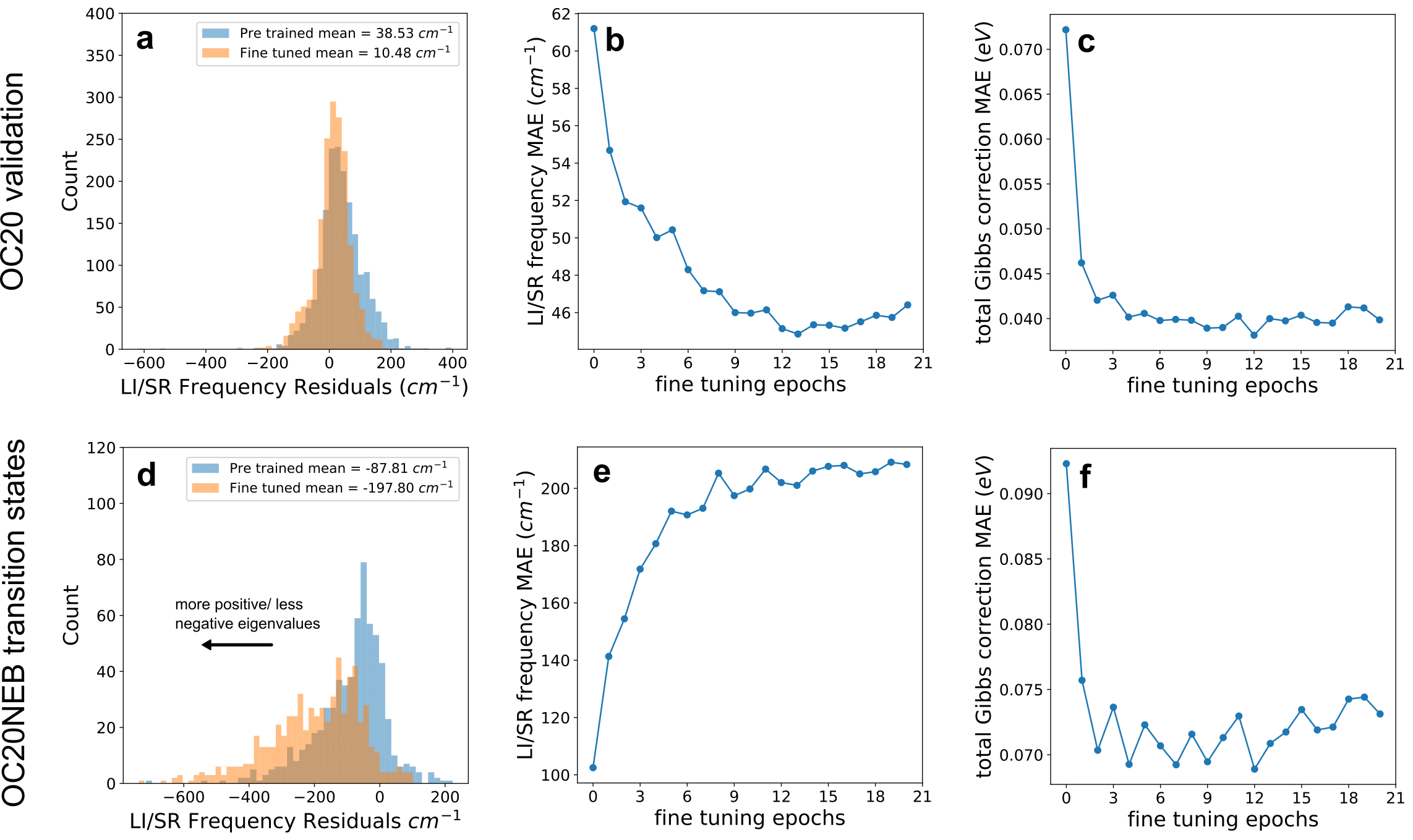}
    \caption{The impact of finetuning for the OC20 validation (top) and OC20NEB transition states (bottom). (a and d) A residual plot of the \gls{LI/SR} frequency for the best performing finetuned checkpoint -- trained for 12 epochs, (b and e) the evolution of the \gls{mae} in the \gls{LI/SR} frequency with number of training epochs, and (c and f) the evolution in the total Gibbs correction \gls{mae} at 300 K with the number of training epochs.}
    \label{fig:finetuning_curves}
\end{figure}

To determine the impact of fine-tuning on model performance for the auxiliary task of determining the numerical Hessian, we fine-tune the model on Hessian singlepoints sampled from local minimums.
The dataset used to train the models during the fine-tuning process is limited to systems that appear in the OC20 training dataset, so transition states are not represented in the data.
For every adsorbate-slab system trained on, two small displacements are included in each direction for every adsorbate atom.
Figure \ref{fig:finetuning_curves} shows that fine-tuning improves the accuracy of predictions of the \gls{LI/SR} frequencies (Fig. \ref{fig:finetuning_curves}b) and the Gibbs free energy corrections for relaxed vibration calculations (OC20 validation) (Fig. \ref{fig:finetuning_curves}c). 
Fine-tuning also improves the Gibbs free energy prediction accuracy for transition state calculations (OC20NEB) (Fig. \ref{fig:finetuning_curves}f), but it degrades the prediction of the \gls{LI/SR} frequencies (Fig. \ref{fig:finetuning_curves}e).
This is sensible since the training data was comprised of non-reactive systems which should not have negative eigenvalues like those of the \gls{LI/SR} frequency of reactive systems.

To test whether training on local minimums explicitly biases the model towards non-complex frequencies, we examine the distribution of residuals for the \gls{LI/SR} transition state frequency between the pretrained and fine-tuned models (after 12 epochs) in Figures \ref{fig:finetuning_curves}a and \ref{fig:finetuning_curves}d.
We observe a shift towards a more normal distribution of residuals in the OC20 validation calculations (Fig. \ref{fig:finetuning_curves}a), while also observing a shift towards a more negatively skewed distribution of residuals for the transition state calculations (Fig. \ref{fig:finetuning_curves}d).
The shift towards negative residuals shows that the lowest frequency predicted by the fine-tuned model is more often erroneously high, shifting towards non-complex frequencies, which would be observed in the training set.
This means that fine-tuning the model on displacements sampled about local minima decreased its propensity to predict imaginary vibrational modes, while increasing its overall predictive accuracy.
Therefore, we conclude fine-tuning on small displacement data around local minima to be an effective strategy for improving the model's Hessian predictions in general, but not for improving its capacity to identify optimal transition states.
This is useful for cases like predicting Gibbs free energy corrections for a wide variety of data sets, including transition state data, where the entire Hessian is used.
However if the Hessian prediction is used for identifying transition states, or transition state optimization strategies, then the training data selected for fine-tuning should be representative of transition state calculations to provide any improvement.

\subsection{Gibbs free energy corrections} 
Using the potential energy numerical Hessians from \gls{dft} and a variety of machine learning models, the Gibbs free energy errors are calculated and compared at 300 K in Table \ref{tab:gibbs_energy_corrections} for the OC20 validation systems (top) and the OC20NEB transition states (bottom). For the OC20 validation systems, \gls{ML} computed approaches may be compared to Adopt-a-value \gls{dft}, which is similar to the approach that is commonly taken by researchers whereby the correction is calculated using \gls{dft} for each adsorbate of interest on a single surface, and then that value is assumed to hold for all surfaces. The Equiformer V2 153M (EqV2 153M) parameter model with a bias correction of 0.240 eV/$\text{\AA}^2$ outperforms the Adopt-a-value approach which have \gls{mae}s of 0.042 eV and 0.058 eV, respectively.

It can be observed that the performance on this auxiliary task is correlated with the OC20 training metrics. EqV2 153M model has the lowest \gls{mae}s for force and energy on OC20 and the remaining ordering is retained. This is somewhat true for the transition states as well, although PaiNN and GemNet T are incorrectly ordered. Across the models, performance is poorer on the transition states than on the OC20 validation systems except SchNet. Poorer performance is expected given that transition states do not appear in the OC20 training data.

Given the functional forms of the \gls{zpe} and entropy terms in Equations \ref{eqn:zpe} and \ref{eqn:ts}, the entropy term is more sensitive to relative errors in smaller frequencies and the \gls{zpe} is more sensitive to relative errors in larger frequencies. For both Adopt-a-value and the EqV2 153M model, the error contributions for these two terms are roughly equal for the OC20 validation set. There is a marked improvement in the \gls{mae} with fine-tuning (fine-tuned EQ2) giving an overall MAE of 0.031 eV for OC20 validation and 0.068 eV for OC20NEB transition states, which is the lowest of all. The percent improvement with fine-tuning compared to the pretrained EqV2 153M model is larger for the ZPE term than the entropy term, indicating that the improvement is not as strong for smaller frequencies.

For the OC20 validation systems, we also present the mean per adsorbate approach. This gives an improvement over Adopt-a-value (0.045 eV  versus 0.058 eV \gls{mae}), but is still bested by the corrected EqV2 153M model. The mean ordered vibrational energies for each adsorbate are included in a table in the Supplementary Information, so that they may be used. The distribution width varies by adsorbate. Sensibly it is roughly correlated with the number of atoms in the adsorbate (larger adsorbates tend to have wider distributions of corrections than smaller adsorbates). A box and whisker plot of the adsorbate Gibbs corrections at 300 K has been included in the Supplementary Information as well.


\begin{table}
\caption{Mean absolute errors of Gibbs energy correction terms. Each correction term contributes to the total Gibbs energy correction. Predictions made on local minimum relaxations and transition state calculations taken from the \gls{oc20} validation set. The methods column corresponds to the \gls{mlp}s used to perform a vibration calculation on each system in the data set, except for the mean per adsorbate approach, which uses the average correction over each adsorbate and only applies to local minimum calculations.}
\label{tab:gibbs_energy_corrections}
\centering

\begin{tabular*}{0.7\linewidth}{@{\extracolsep{\fill}} lrrrr}
\toprule 
\multicolumn{5}{c}{\textbf{Local Minimum Correction MAE}} \\ 
\midrule
 Method             &   ZPE &    TS &   C$\mathrm{_p}$ &   Total Gibbs \\
                    & $[eV]$& $[eV]$&            $[eV]$&         $[eV]$\\
\midrule
 
 Fine-tuned EQ2 153M    & 0.017 & 0.029 &            0.010 &         0.031 \\
 EQ2 153M corrected     & 0.025 & 0.035 &            0.012 &         0.042 \\
 Mean per adsorbate DFT & 0.029 & 0.035 &            0.012 &         0.045 \\
 Adopt-a-value DFT      & 0.042 & 0.040 &            0.015 &         0.058 \\
 EQ2 153M               & 0.042 & 0.041 &            0.015 &         0.062 \\
 EQ2 31M                & 0.059 & 0.047 &            0.017 &         0.080 \\
 GemNet T               & 0.074 & 0.054 &            0.020 &         0.101 \\
 PaiNN                  & 0.121 & 0.066 &            0.023 &         0.151 \\
 SchNet                 & 0.147 & 0.087 &            0.030 &         0.201 \\
\bottomrule
\end{tabular*}
 
\begin{tabular*}{0.7\linewidth}{@{\extracolsep{\fill}} lrrrr}
\toprule 
\multicolumn{5}{c}{\textbf{Transition State Correction MAE}} \\ 
\midrule
 Method         &   ZPE &    TS &   C$\mathrm{_p}$ &   Total Gibbs \\
                & $[eV]$& $[eV]$&            $[eV]$&         $[eV]$\\
\midrule
 EQ2 153M corrected  & 0.046 & 0.035 &            0.015 &         0.064 \\
 Fine-tuned EQ2 153M & 0.047 & 0.038 &            0.016 &         0.068 \\
 EQ2 153M            & 0.066 & 0.048 &            0.020 &         0.092 \\
 GemNet T            & 0.071 & 0.053 &            0.021 &         0.102 \\
 EQ2 31M             & 0.098 & 0.070 &            0.029 &         0.139 \\
 SchNet              & 0.108 & 0.065 &            0.027 &         0.143 \\
 PaiNN               & 0.158 & 0.098 &            0.036 &         0.217 \\
\bottomrule
\end{tabular*}

\end{table}

\subsection{Complete potential energy sampling} 
\begin{figure}
    \centering
    \includegraphics[width=1.0\linewidth]{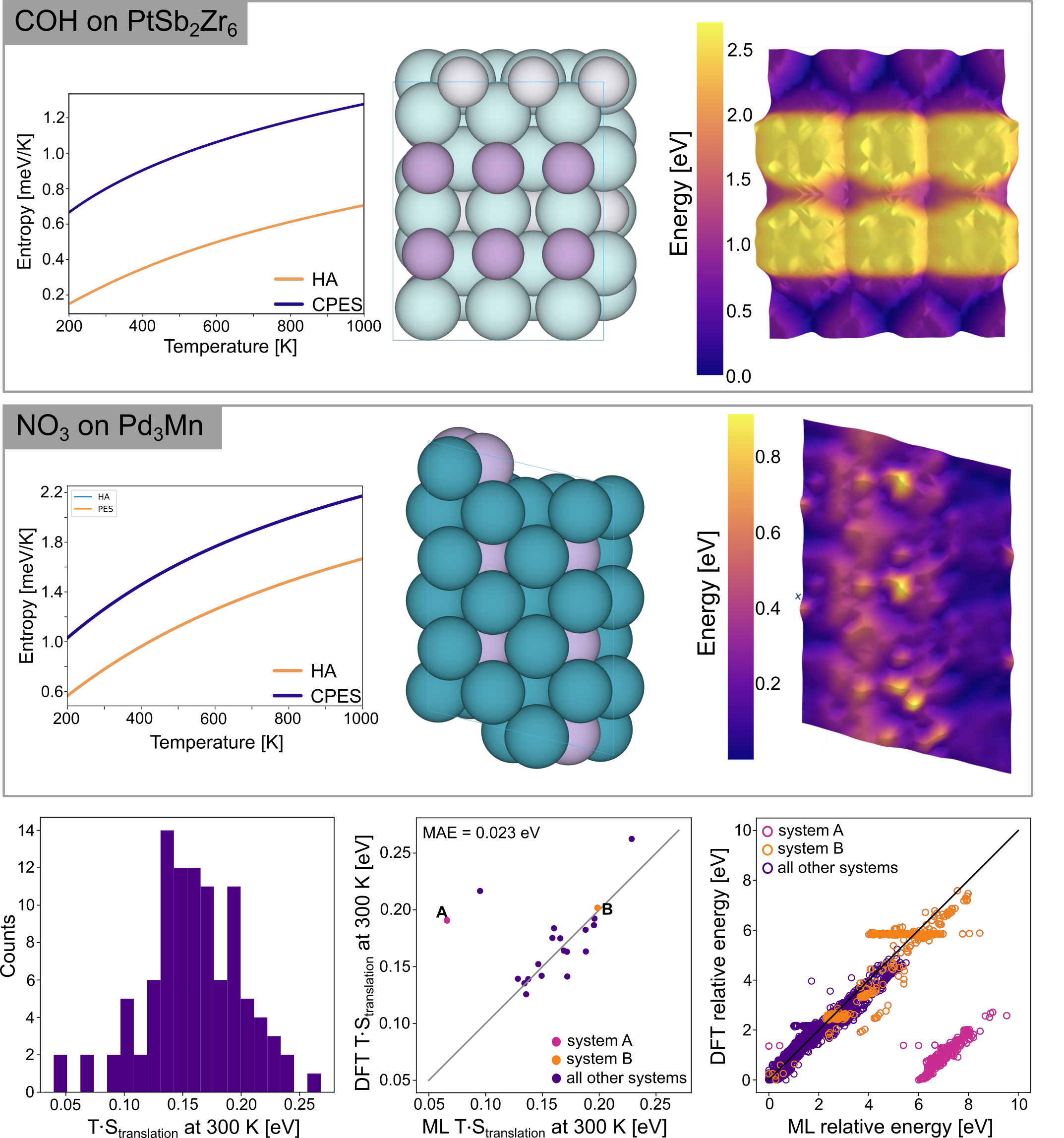}
    \caption{Two examples of the complete potential energy sampling (CPES) technique for COH on an Pt-Sb-Zr alloy and NO$_3$ on an Pd-Mn alloy. For each, the slab (center), potential energy surface (right), and the entropy profiles versus the temperature is shown for the harmonic approximation (HA) and CPES (left). The distribution of the translational component to the entropy correction evaluated at 300 K across 100 randomly sampled systems from validation (bottom, left). The parity between \gls{ML} determined and \gls{dft} single point validated  translational entropy Gibbs corrections (bottom, center). The parity between the relative adsorption energies used to construct the PES from \gls{ML} and \gls{dft} single points (bottom, right). In the bottom, center and bottom, right figures two systems (A and B) have been highlighted so the impact of outliers in the relative energies (right) on the translational entropy term (center) can be seen.}
    \label{fig:cpes}
\end{figure}

We use the Equiformer V2 31M parameter model to perform complete potential energy sampling (CPES) to obtain a more complete entropy correction term of the Gibbs free energy for 100 systems from the 1,812 OC20 validation set. A summary of the results from this analysis is shown in Figure \ref{fig:cpes}. For two example systems, the temperature profile of the entropy using the harmonic approximation and adding the \gls{cpes} translational entropy are shown (left), along with an image of the slab (center), and the potential energy surface (PES). For both examples, there are large swaths of accessible (low change in energy) states, which can be seen in the \gls{PES}. For the top example, the Sb (purple) atoms are unfavorable for COH to bind to, which gives the high energy (yellow) regions on the \gls{PES}.

Across all 100 systems, the average difference in the correction with \gls{cpes} was 0.16 eV at 300 K, with a maximum of 0.27 eV. The distribution of all differences in the correction are shown in the lower left of Figure \ref{fig:cpes}.  There were some additional issues with errors in the models because of the Hookean and fixed center of mass constraints. This resulted in a loss of density for the meshgrid, which for most adsorbate-surface combinations was modest (4\% of placements on average), but in some cases caused the approach to be unusable (a maximum of 48\% was observed). For adsorbates that are stable in the gas phase, there were some adsorbate-surface combinations where a significant number of adsorbate placements desorbed over relaxation. This undermines the fidelity of the approach which is intended to capture 2D translational freedom, not 3D.

For a subset of 17 systems which did not have a significant proportion of desorbed or errant systems, DFT single points were performed on the \gls{ML} relaxed structures. A comparison of the relative DFT single point energies (relative to the minimum) and the relative ML energies can be seen in the lower, right of Figure \ref{fig:cpes}. The majority of systems have good energetic parity, but there are a few exceptions. For system B (orange), there are a large number of points which are determined to be equal in energy by DFT, but ML erroneously finds the energies to have a distribution of values. For system A, which is shown in pink, the minimum energy by is erroneously small and far away from the distribution. This is the reason that the primary cluster of data is shifted away from the parity line but still correlated. The impact these artifacts have on the translational entropy component to the Gibbs free energy is shown in the bottom, center of Figure \ref{fig:cpes}. The differences with system B have a minute impact on the overall correction, while the differences with system A have a large impact. Similar artifacts can be seen for the OC20Dense dataset\cite{lan2023adsorbml} for the GemNet-OC model\cite{gasteiger2022gemnet}, and a figure supporting this has been included in the Supplementary Information. There are two large outliers when comparing the translational entropy terms by ML and supplemented with DFT single points. Both of these were caused by the same issue described for system A: having a few ML energies that are far away from the energy distribution. The other outlier can not be seen the bottom, right of Figure \ref{fig:cpes} because the overall magnitude of its energy distribution is small so it is masked by the noise in other systems. In these cases, the translational entropy has been underestimated by ML. Still, the \gls{mae} in the translational entropy term by ML with reference to the DFT single points is just 0.023 eV at 300 K.

The results from J{\o}rgensen et al.\cite{jorgensen2017Adsorbate} for CO and O on the Pt (111) surface were reproduced with \gls{ML} and \gls{dft} relaxations. A comparison of the approaches reveals very good agreement and a figure supporting this has been included in the Supplementary Information.

Overall there is good agreement between the \gls{dft} and \gls{ML} \gls{PES} and Gibbs corrections, but the outlier in the bottom, center figure suggests that supplementing with DFT single points can be advantageous. These results underscore the importance of considering translational degrees of freedom for the entropy in adsorbed systems. The translational entropy correction will be even more impactful at higher temperatures and therefore should not be ignored. The \gls{mlp}s perform well for this task, which would be very expensive to perform with \gls{dft}. 

\subsection{Transition state search with GNNs}

\begin{table}
\small
\caption{A comparison of key metrics for the CatTSunami baseline and 2 applications of Sella for transition state optimization. Sella refined (any converged) is the case where we take converged Sella results but fall back on converged \gls{neb}s when Sella does not converge, and Sella refined (both converged) is the case where we only keep results where both the \gls{neb} and Sella converge.} 
\label{tab:sella_results}
\centering

\begin{tabular}{lrrrr}
\toprule
 Method & Mean fmax  & Energy MAE  & Converged  & Success \\
        & [eV/Ang]   & [eV]        & [\%]       & [\%]  \\    
\midrule
CatTSunami baseline & 0.159 (0.078) & 0.060 (0.149) & 80.15 & 91.54 \\
Sella refined (any converged) & 0.137 (0.074) & 0.069 (0.186) & 92.99 & 89.79 \\
Sella refined (both converged) & 0.132 (0.073) & 0.041 (0.113) & 70.60 & 95.20 \\
\bottomrule
\end{tabular}

\end{table}

We use Sella to perform transition state optimization on the \gls{ML}-relaxed transition states from the OC20NEB\cite{wander2024cattsunami} dataset.
Table \ref{tab:sella_results} shows that depending how it is applied, Sella may be used to improve either the success or convergence rate depending on the user's priority for these outcomes.
By selecting only systems where both the \gls{neb} algorithm and the subsequent Sella algorithm converged more accurate energies (0.041 eV MAE), lower forces (0.132 mean \gls{Fmax}), and a higher success rate (95.2\%) are achieved. 
95\% success can be seen as the maximum success rate because this is the success rate obtained in the CatTSunami baseline when \gls{ML} is used to pre-optimize the \gls{neb}, followed by a \gls{dft} \gls{neb}. 
By this approach, despite the \gls{neb} being constrained about the reactant and product configurations, there are 5\% of energetically dissimilar cases due to differences in the initialization point of the intermediate frames.
These improvements come with the cost of accepting a smaller proportion of systems (9.6\% fewer converged).
This is a fairly significant trade off when converging as many systems as possible is a priority, but it does provide higher confidence in the accuracy of the calculations which are accepted.
Alternatively the Sella refinement can be used on all systems, and any systems where either the \gls{neb} or Sella algorithm converges are accepted to improve overall convergence (93\%) when compared to using the \gls{neb} alone (80.2\% converged).
In this case the proportion of systems which converge is greatly improved when compared to the CatTSunami baseline (93.0\% versus 80.2\%), while the energy \gls{mae} is only marginally degraded (0.069 eV versus 0.060 eV MAE).
The max force also improves, indicating that the Sella algorithm may be more successful than the \gls{neb} algorithm at finding valid relaxed saddle points with the \gls{mlp} according to \gls{dft}.

\begin{figure}
    \centering
    \includegraphics[width=1.0\linewidth]{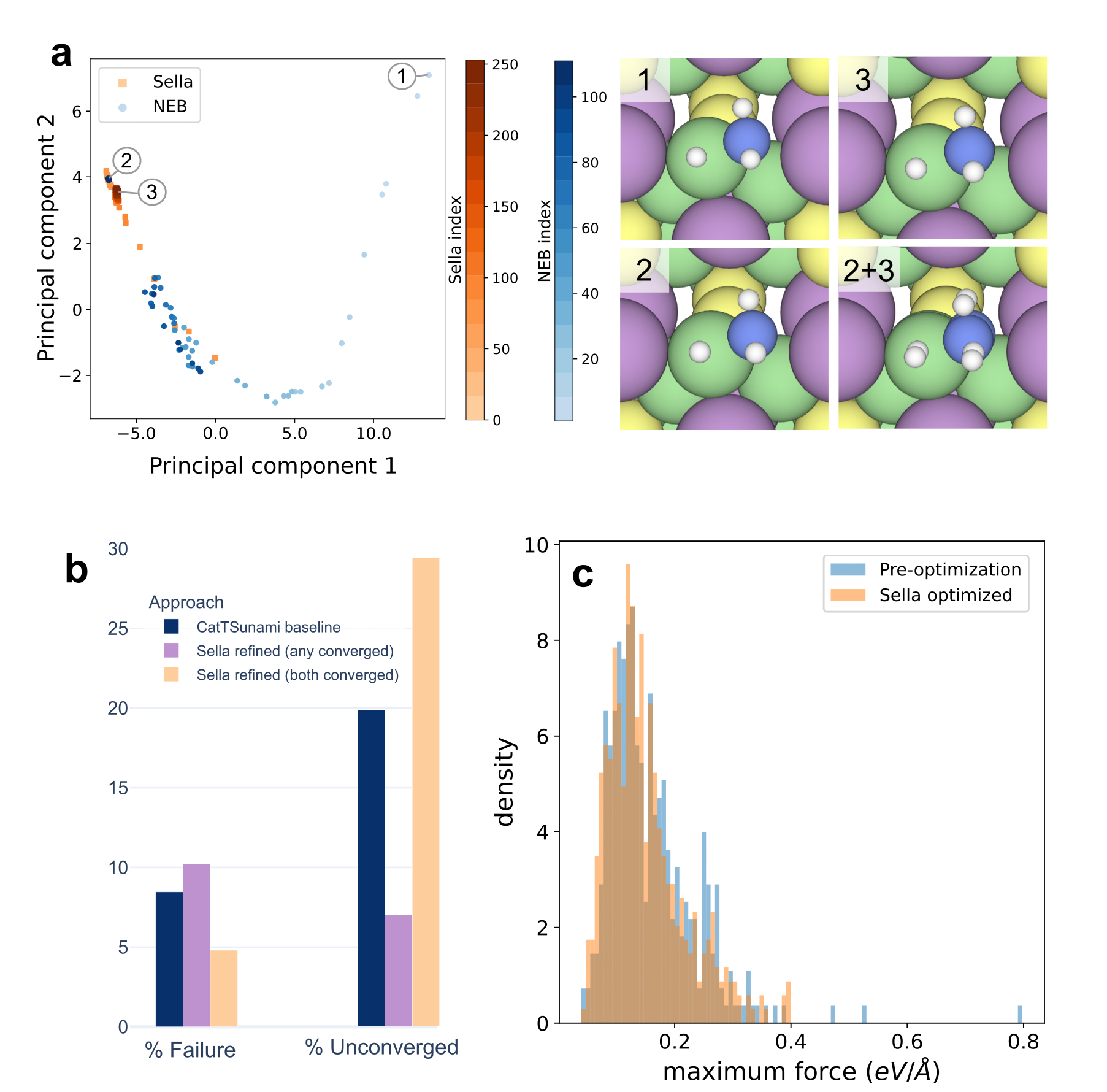}
    \caption{
    A summary of results for using Sella\cite{hermes2022sella} to perform transition state optimization on OC20NEB systems. 
    (a) A look at the evolution of one system across the \gls{neb} trajectory and the Sella optimization trajectory using principal component analysis. 
    The unrelaxed, \gls{neb} relaxed, and Sella refined transition states are shown in images 1, 2, and 3, respectively. 
    (b) The \% failures and \% uncoverged through the use of Sella refinement by two different approaches. 
    (c) The distribution of the maximum force (by DFT single point) before and after Sella refinement.
    }
    \label{fig:sella}
\end{figure}

The percent of systems that are unsuccessful and unconverged are shown in Figure \ref{fig:sella}b. 
Using Sella to refine with any converged to optimize for convergence makes a significant difference, reduces the number of unconverged systems by 65\% over the baseline.
Using Sella to refine with both converged reduces the number of unsuccessful systems by 41\%.
A benefit of using the Sella refinement on top of the \gls{neb} algorithm can be seen in Fig. \ref{fig:sella}c, where the maximum force distribution is plotted before and after Sella refinement.
When the Sella algorithm converges, it tends to remove higher force outliers from the distribution, resulting in more structures which are reasonably close to convergence.
It also results in a modest overall shift of the data towards lower \gls{Fmax} values.
In Figure \ref{fig:sella}a we see an example of what this refinement does to the transition state structure.
Plotted on the left are the first two principal components of a principal component analysis performed on the latent space of Equiformer V2, according to the approach used by Musielewicz et al.\cite{musielewicz2024improveduncertaintyestimationgraph}.
We observe that the Sella optimization retraces a similar trajectory to the end of the \gls{neb} optimization, and hones in on a slightly more relaxed transition state.
This results in a subtle difference in positions of the adsorbate atoms, as seen on the right (2+3).
In practice, combining these transition state optimizations, with \gls{mlp} driven estimates of Gibbs free energy corrections, promises to enable \gls{mlp}s to make wholesale predictions of key kinetic parameters.

\section{Conclusion}
In this work, we demonstrate that models pretrained on the OC20 dataset\cite{chanussot2021open} are able to determine numerical energy Hessians for in domain systems with good accuracy. To facilitate the assessment of model performance, we constructed a dataset of 4,599 numerical revised Perdew-Burke-Ernzerhof (RPBE) \gls{dft} Hessians. In our assessment, we considered four different pretrained model architectures, and also considered the opportunity to fine-tune a model to improve performance.

To understand the implications of this auxiliary ability, we considered the usefulness in application of the energy Hessian to Gibbs free energy corrections and transition state optimization. For Gibbs free energy corrections, the top performing model has an \gls{mae} that outperforms (0.042 versus 0.058 eV at 300 K) the commonly used practice of calculating the per adsorbate correction on one surface and taking that value to be true for all other surfaces. We also explored the opportunity to utilize pretrained \gls{mlp}s to treat translational entropy in the Gibbs free energy corrections, going beyond the harmonic approximation. We found that 94\% of systems had a translational entropy contribution greater than 0.1 eV at 300 K. This effect, of course, would be more pronounced at higher temperatures. For transition state optimization using the Hessian we utilized Sella\cite{hermes2022sella} with the \gls{mlp}s to refine the ML-relaxed transition states from the OC20NEB dataset. We found that Sella can be used to increase the convergence rate from 80.2\% to 93.0\%, a 65\% reduction in the number of unconverged systems.

\begin{acknowledgement}
We would like to thank Brandon Wood for running some of the DFT calculations used in this work. We would like to thank Samuel Blau for his guidance. This research used resources of the National Energy Research Scientific Computing Center (NERSC), a U.S. Department of Energy Office of Science User Facility located at Lawrence Berkeley National Laboratory, operated under Contract No. DE-AC02-05CH11231 using NERSC award BES-ERCAPm2755. R. Cheula acknowledges support from European Union’s Horizon Europe research and innovation programme under the Marie Skłodowska-Curie grant no. 101108769.
\end{acknowledgement}

\section{Contributions}
\begin{enumerate}

\item Brook Wander: conceptualization, data curation, formal analysis, investigation, software, writing -- original draft, writing -- review \& editing, validation, visualization
\item Joseph Musielewicz: conceptualization, software, investigation, methodology, visualization, writing -- original draft, writing -- review \& editing
\item Raffaele Chuela: software, conceptualization, methodology, writing -- review \& editing, writing -- original drafts
\item John R. Kitchin: writing -- review \& editing, conceptualization, funding acquisition
\end{enumerate}
\clearpage
\bibliography{bib}
\clearpage

\section{Supplementary Information}
\subsection{DFT settings}
DFT calculations were performed with \gls{vasp}\cite{kresse1994ab, kresse1996efficiency, kresse1996efficient, kresse1999ultrasoft} with periodic boundary conditions and the projector augmented wave (PAW) pseudopotentials \cite{kresse1999ultrasoft, blochl1994projector}. The external electrons were expanded in plane waves with kinetic energy cut-offs of 350 eV. Exchange and correlation effects were taken into account via the generalized gradient approximation and the revised Perdew-Burke-Ernzerhof (RPBE)\cite{perdew1996generalized, zhang1998comment} functional, because of its improved description of the energetics of atomic and molecular bonding to surfaces\cite{hammer1999improved}. Bulk and surface calculations were performed considering a K-point mesh for the Brillouin zone derived from the unit cell parameters as an on-the-spot method, employing the Monkhorst-Pack grid\cite{monkhorst1976special}.

\subsection{Sella hyperparameters}
\begin{table}
\caption{The optimized Sella hyperparameters, for application to OC20-like systems using the Equiformer V2 153M parameter model.}
\label{tab:sella_hyp}
\centering

\begin{tabular}{lrr}
\toprule 
Parameter & Value & Description \\
\midrule

$\gamma$ & 0 & Convergence criterion for iterative diagonalization. \\
$\eta$ & 7.0E-4 & Finite difference step size ($\text{\AA}$). \\
$\delta_0$ & 4.8E-2 & Initial trust radius ($\text{\AA}$). \\
$\rho_{inc}$ & 1.035 & Threshold for increasing trust radius. \\
$\rho_{dec}$ & 5.0 & Threshold for decreasing trust radius. \\
$\sigma_{inc}$ & 1.15 & Trust radius increase factor. \\
$\sigma_{dec}$ & 0.65 & Trust radius decrease factor. \\
method & P-RFO & Choice of optimization algorithm. \\

\bottomrule
\end{tabular}

\end{table}

\subsection{Per model Hessian performance}
\begin{tabular}{lrrr}
\toprule 
\multicolumn{4}{c}{\textbf{Local Minimum MAEs}} \\ 
\midrule
 Method         &   Eigenvalues &   All Frequencies &   LI/SR Frequencies \\
                &     $eV/\text{\AA}^2$&          $cm^{-1}$&            $cm^{-1}$\\
\midrule
 Fine tuned EQ2 &         0.513 &            29.369 &              46.446 \\
 EQ2 153M       &         1.485 &            58.195 &              63.588 \\
 EQ2 31M        &         2.155 &            80.027 &              79.687 \\
 GemNet T       &         2.580 &           101.830 &              95.677 \\
 PaiNN          &         4.495 &           165.483 &             140.570 \\
 SchNet         &         5.779 &           193.191 &             115.808 \\
\bottomrule
\end{tabular}

\begin{tabular}{lrrr}
\toprule 
\multicolumn{4}{c}{\textbf{Transition State MAEs}} \\ 
\midrule
 Method         &   Eigenvalues &   All Frequencies &   LI/SR Frequencies \\
                &     $eV/\text{\AA}^2$&          $cm^{-1}$&            $cm^{-1}$\\
\midrule
 Fine tuned EQ2 &         1.191 &            74.549 &             202.010 \\
 EQ2 153M       &         1.797 &            88.067 &             102.550 \\
 GemNet T       &         1.950 &            96.581 &             109.430 \\
 EQ2 31M        &         2.785 &           136.198 &             216.724 \\
 SchNet         &         3.287 &           163.642 &             238.127 \\
 PaiNN          &         4.139 &           217.474 &             264.510 \\
\bottomrule
\end{tabular}

\subsection{Vibrational correction distributions}
\begin{figure}
    \centering
    \includegraphics[width=0.95\linewidth]{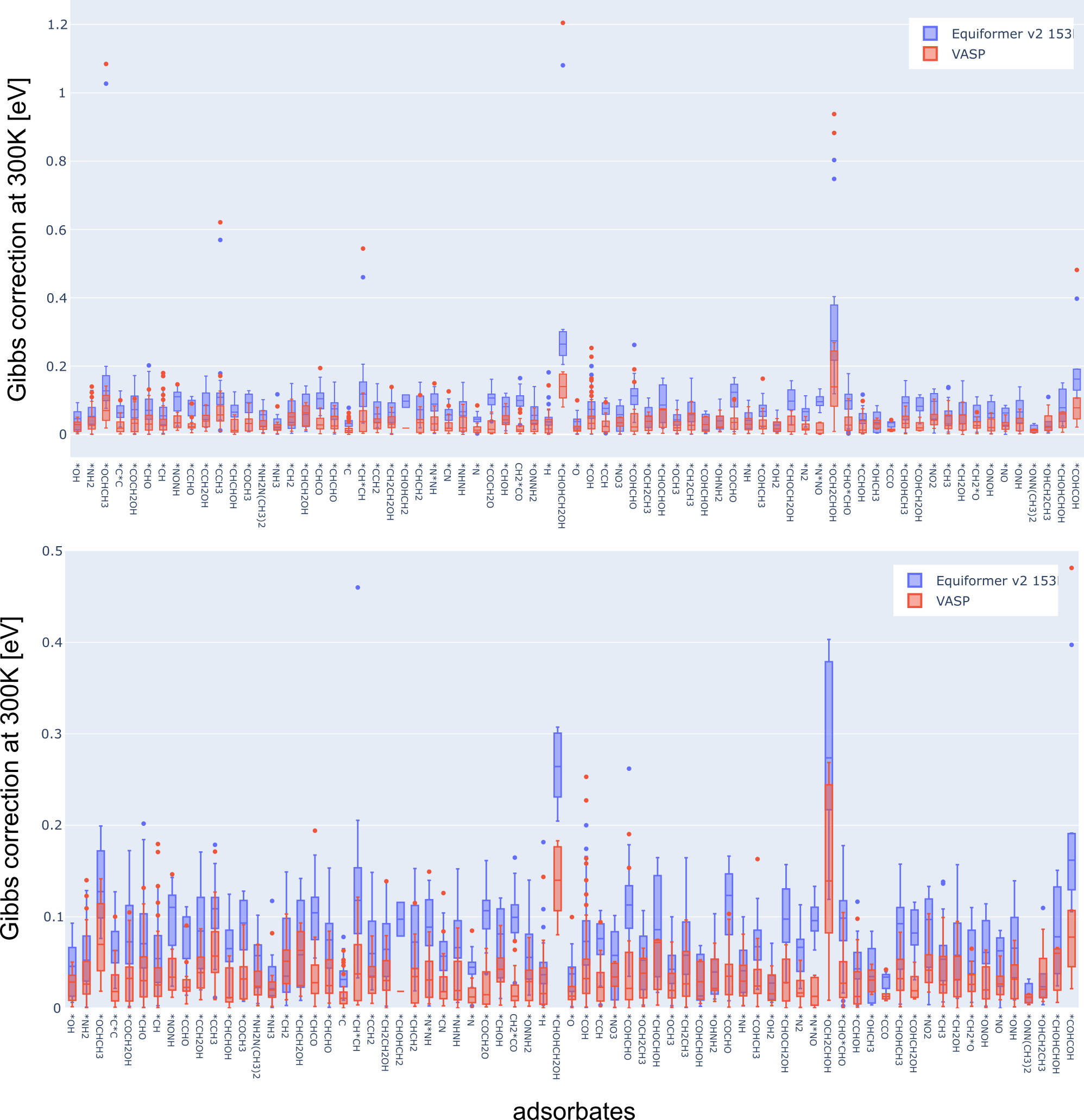}
    \caption{
    The distribution of Gibbs corrections per adsorbate at 300 K.
    }
    \label{fig:distribution-gibbs-2}
\end{figure}

\subsection{OC20Dense single point parities}
\begin{figure}
    \centering
    \includegraphics[width=0.95\linewidth]{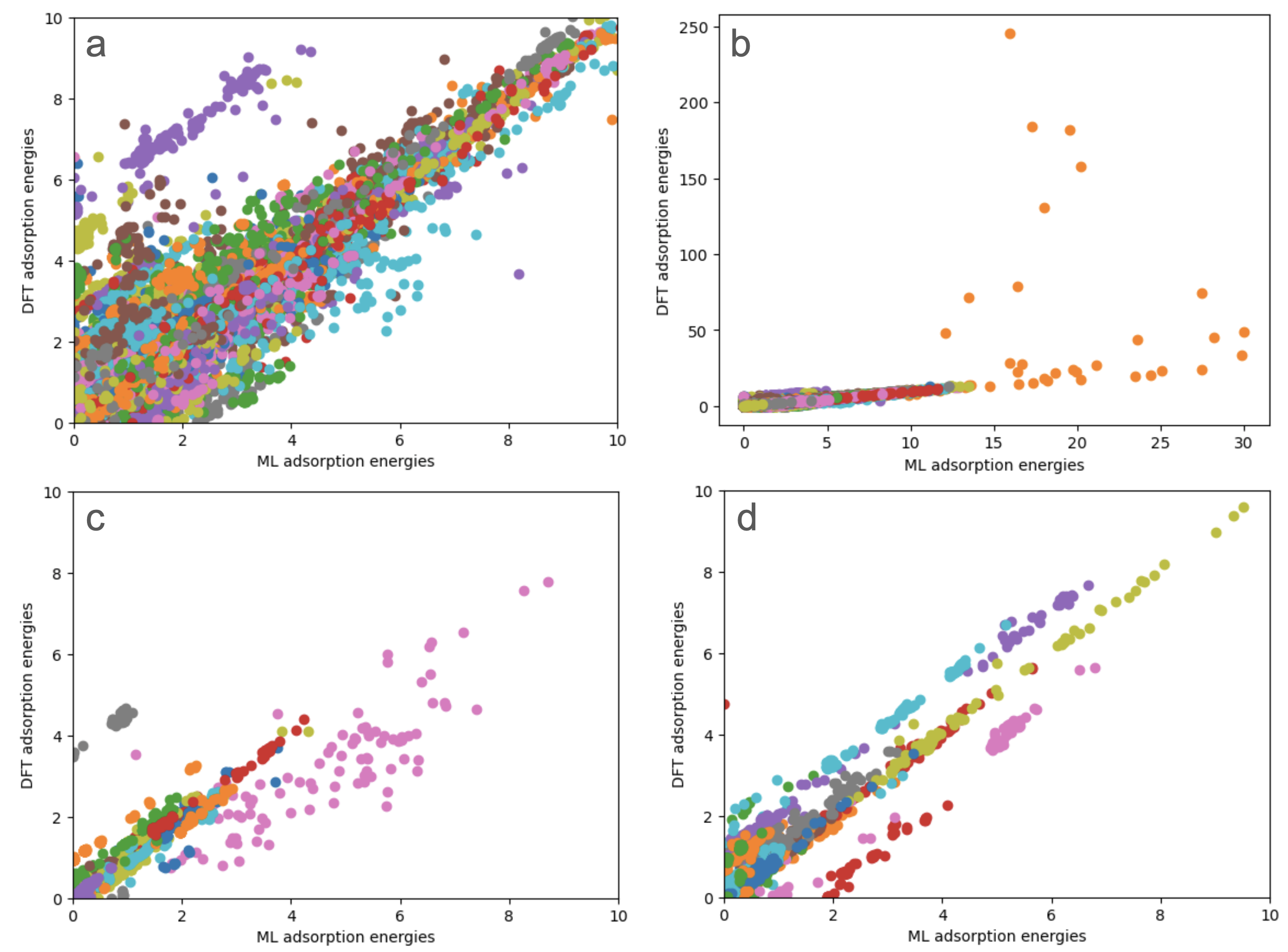}
    \caption{A plot of the relative adsorption energy (relative to the minimum energy) for DFT single points (y-axis) and GemNet-oc ML energies (x-axis) for (a) all OC20Dense systems,  (b) all OC20Dense systems with the plot constrained from 0 to 10 eV, and (c,d) A random subselection of systems so aritfacts may be seen in more detail. Each color is an adsorbate + surface combination, though for the plots that show all data, colors are not unique to a single system.
    }
    \label{fig:distribution-gibbs}
\end{figure}

\subsection{Mean frequencies per adsorbate}

{\footnotesize
\begin{longtable}{p{2.5cm}p{13.5cm}}
\toprule
adsorbate & mean frequencies \\
\midrule
*OH & [108, 180, 276, 447, 718, 3623] \\
*NH2 & [133, 236, 364, 483, 637, 720, 1488, 3273, 3384] \\
*OCHCH3 & [26, 69, 107, 153, 176, 224, 279, 473, 710, 841, 954, 1029, 1172, 1272, 1321, 1366, 1457, 2717, 2797, 2850, 2909] \\
*C*C & [225, 292, 357, 419, 567, 1490] \\
*COCH2OH & [35, 75, 103, 143, 190, 259, 326, 408, 490, 570, 866, 956, 1033, 1140, 1234, 1305, 1372, 1496, 2875, 2977, 3476] \\
*CHO & [114, 185, 251, 332, 472, 659, 1108, 1325, 2694] \\
*CH & [244, 367, 494, 634, 771, 2911] \\
*NONH & [45, 111, 157, 217, 288, 340, 619, 745, 1001, 1230, 1402, 3194] \\
*CCHO & [72, 145, 199, 256, 314, 417, 665, 860, 1107, 1261, 1428, 2842] \\
*CCH2OH & [40, 94, 167, 214, 280, 345, 436, 585, 830, 939, 1041, 1130, 1220, 1310, 1422, 2784, 2898, 3585] \\
*CCH3 & [55, 113, 179, 271, 324, 477, 881, 933, 1016, 1272, 1362, 1375, 2755, 2819, 2861] \\
*CHCHOH & [43, 77, 120, 163, 290, 401, 562, 630, 710, 890, 992, 1130, 1264, 1319, 1507, 2947, 3058, 3316] \\
*COCH3 & [28, 93, 124, 159, 235, 273, 402, 556, 872, 940, 1044, 1263, 1361, 1404, 1532, 2830, 2988, 3031] \\
*NH2N(CH3)2 & [10, 32, 58, 85, 107, 163, 244, 270, 348, 412, 473, 540, 800, 971, 1011, 1072, 1090, 1121, 1199, 1235, 1301, 1388, 1413, 1425, 1433, 1444, 1454, 1570, 2844, 2861, 2967, 2982, 3025, 3036, 3143, 3300] \\
*NH3 & [61, 121, 168, 285, 516, 546, 1175, 1571, 1586, 3143, 3359, 3402] \\
*CH2 & [123, 277, 385, 487, 675, 765, 1315, 2820, 2959] \\
*CHCH2OH & [28, 92, 148, 185, 290, 386, 468, 583, 728, 896, 944, 1067, 1146, 1205, 1297, 1350, 1417, 2828, 2930, 2983, 3495] \\
*CHCO & [74, 122, 187, 236, 332, 411, 535, 641, 887, 1105, 1682, 2899] \\
*CHCHO & [70, 112, 169, 242, 315, 387, 635, 744, 901, 1015, 1192, 1295, 1458, 2842, 2979] \\
*C & [290, 413, 634] \\
*CH*CH & [144, 238, 273, 395, 491, 586, 799, 920, 1096, 1297, 2812, 2922] \\
*CCH2 & [101, 175, 267, 325, 414, 497, 827, 946, 1285, 1488, 2850, 3022] \\
*CH2CH2OH & [21, 70, 104, 145, 220, 327, 397, 496, 673, 857, 923, 1010, 1088, 1151, 1224, 1288, 1359, 1390, 1442, 2862, 2920, 2955, 3004, 3487] \\
*CHOHCH2 & [96, 142, 191, 243, 339, 409, 468, 542, 709, 882, 1010, 1054, 1077, 1131, 1211, 1354, 1379, 2940, 2973, 3044, 3548] \\
*CHCH2 & [110, 173, 244, 319, 437, 525, 769, 892, 974, 1186, 1284, 1466, 2685, 2923, 3032] \\
*N*NH & [143, 231, 289, 388, 458, 599, 1031, 1294, 3184] \\
*CN & [127, 209, 315, 358, 438, 1699] \\
*NHNH & [96, 151, 222, 293, 376, 480, 802, 1037, 1256, 1389, 3185, 3248] \\
*N & [228, 351, 612] \\
*COCH2O & [59, 107, 155, 198, 278, 339, 401, 503, 547, 832, 957, 1032, 1168, 1257, 1342, 1471, 2844, 2921] \\
*CHOH & [64, 143, 218, 305, 435, 501, 842, 1056, 1194, 1339, 2754, 3359] \\
CH2*CO & [91, 151, 202, 269, 332, 428, 568, 659, 776, 926, 1068, 1307, 1485, 2971, 3100] \\
*ONNH2 & [16, 38, 73, 116, 145, 182, 376, 584, 692, 1020, 1189, 1326, 1527, 3164, 3475] \\
*H & [455, 714, 1468] \\
*CHOHCH2OH & [45, 65, 93, 139, 178, 217, 277, 358, 409, 494, 592, 768, 821, 921, 959, 1016, 1072, 1137, 1219, 1270, 1307, 1364, 2699, 2759, 2789, 3202, 3301] \\
*O & [179, 288, 518] \\
*COH & [76, 162, 248, 312, 401, 511, 996, 1183, 3352] \\
*CCH & [112, 212, 291, 355, 440, 615, 815, 1534, 3130] \\
*NO3 & [34, 60, 79, 137, 187, 243, 603, 623, 719, 890, 1108, 1277] \\
*COHCHO & [55, 113, 152, 197, 235, 294, 362, 415, 480, 595, 810, 965, 1068, 1183, 1295, 1443, 2907, 3495] \\
*OCH2CH3 & [13, 43, 79, 127, 179, 232, 311, 477, 786, 876, 1020, 1076, 1122, 1254, 1332, 1356, 1431, 1438, 1455, 2884, 2917, 2944, 3013, 3028] \\
*CHOCHOH & [52, 93, 136, 175, 283, 319, 342, 483, 579, 662, 781, 907, 1001, 1097, 1184, 1264, 1321, 1429, 2922, 3084, 3514] \\
*OCH3 & [42, 81, 118, 169, 244, 373, 1027, 1124, 1141, 1414, 1433, 1440, 2887, 2954, 2973] \\
*CH2CH3 & [36, 88, 134, 204, 243, 400, 577, 899, 933, 1003, 1146, 1213, 1337, 1383, 1430, 1448, 2773, 2870, 2937, 2980, 3014] \\
*COHCHOH & [6, 38, 69, 136, 196, 231, 316, 392, 435, 488, 568, 800, 996, 1093, 1187, 1269, 1318, 1549, 3032, 3359, 3530] \\
*OHNH2 & [18, 68, 114, 142, 197, 265, 548, 785, 1132, 1249, 1326, 1578, 3166, 3297, 3437] \\
*COCHO & [90, 131, 184, 242, 285, 340, 392, 485, 593, 785, 1009, 1148, 1296, 1457, 2972] \\
*NH & [194, 305, 482, 585, 783, 3331] \\
*COHCH3 & [18, 54, 85, 135, 220, 250, 347, 503, 640, 866, 961, 1031, 1156, 1277, 1339, 1392, 1431, 2798, 2961, 3002, 3329] \\
*OH2 & [52, 115, 178, 284, 470, 549, 1547, 3287, 3538] \\
*CHOCH2OH & [50, 85, 121, 153, 196, 265, 323, 411, 513, 610, 787, 872, 966, 1024, 1112, 1185, 1259, 1315, 1349, 1457, 2742, 2914, 2985, 3479] \\
*N2 & [143, 201, 319, 360, 398, 1553] \\
*N*NO & [67, 128, 184, 269, 347, 439, 756, 1114, 1419] \\
*OCH2CHOH & [29, 71, 115, 158, 191, 234, 281, 342, 390, 478, 645, 792, 875, 945, 1006, 1079, 1128, 1201, 1276, 1359, 2462, 2579, 2782, 3311] \\
*CHO*CHO & [77, 113, 153, 198, 256, 307, 398, 512, 718, 774, 837, 952, 1118, 1230, 1337, 1450, 2884, 2955] \\
*CCHOH & [24, 83, 148, 215, 291, 357, 582, 644, 846, 1053, 1197, 1297, 1479, 3016, 3427] \\
*OHCH3 & [14, 55, 91, 138, 162, 248, 537, 957, 1064, 1136, 1290, 1415, 1435, 1449, 2843, 3006, 3049, 3298] \\
*CCO & [33, 63, 184, 265, 335, 529, 577, 1243, 2044] \\
*CHOHCH3 & [48, 85, 115, 194, 228, 269, 371, 432, 508, 791, 930, 997, 1046, 1119, 1223, 1315, 1354, 1418, 1451, 2833, 2894, 2976, 3011, 3440] \\
*COHCH2OH & [33, 68, 103, 137, 196, 240, 291, 363, 462, 505, 595, 874, 952, 999, 1083, 1151, 1236, 1296, 1333, 1423, 2887, 2944, 3315, 3504] \\
*NO2 & [94, 134, 187, 258, 302, 401, 621, 864, 1055] \\
*CH3 & [88, 163, 212, 420, 601, 659, 1172, 1359, 1391, 2782, 2937, 2994] \\
*CH2OH & [34, 88, 136, 240, 354, 491, 642, 896, 1079, 1161, 1284, 1410, 2921, 3016, 3530] \\
*CH2*O & [85, 144, 200, 295, 376, 520, 908, 1132, 1217, 1445, 2829, 2943] \\
*ONOH & [28, 76, 116, 144, 181, 225, 468, 599, 831, 1084, 1361, 3440] \\
*NO & [68, 123, 257, 321, 445, 1377] \\
*ONH & [91, 148, 200, 278, 364, 531, 963, 1325, 3157] \\
*ONN(CH3)2 & [9, 22, 36, 60, 77, 106, 127, 150, 212, 305, 349, 403, 665, 824, 1000, 1028, 1065, 1102, 1185, 1272, 1304, 1384, 1409, 1421, 1431, 1446, 1464, 2932, 2945, 2998, 3015, 3060, 3069] \\
*OHCH2CH3 & [8, 22, 37, 82, 96, 150, 227, 336, 460, 798, 851, 1007, 1038, 1108, 1231, 1285, 1356, 1379, 1438, 1445, 1462, 2912, 2942, 2982, 3015, 3028, 3519] \\
*CHOHCHOH & [17, 69, 106, 139, 185, 232, 298, 352, 447, 553, 634, 734, 876, 1026, 1083, 1154, 1251, 1275, 1329, 1523, 2921, 3058, 3245, 3524] \\
*COHCOH & [52, 81, 119, 194, 226, 247, 286, 359, 466, 534, 701, 889, 1023, 1137, 1237, 1344, 3176, 3345] \\
\bottomrule
\bottomrule
\end{longtable}}

\subsection{Numerical Differentiation vs Analytical Differentiation}
Unlike \gls{vasp}, the outputs of \gls{gnn} potentials can be differentiated analytically. 
This allows the the Hessian matrix to be computed from the first derivative of the predictions analytically, instead of by finite differences.
We implemented both approaches, and found the predicted Hessian to achieve close agreement ($R^2$ = 0.995) between the two approaches for a specific finite difference displacement size of 1.0E-3 \AA{}.
This differs slightly from the displacement size of 7.5E-3 actually used for the \gls{vasp} Hessian dataset, which introduced some noisy error when computing the agreement between the \gls{vasp} Hessian and the analytical \gls{gnn} Hessian.
This error was not produced when using the numerical \gls{gnn} Hessian produced by finite differences with the same displacement size as \gls{vasp}.
Analyzing the runtimes for both the numerical and analytical approaches of obtaining potential energy Hessian with the highest performing \gls{gnn} architectures, we observed that our implementation of automatic differentiation (using the autograd package) required significantly longer compute times to resolve.
On average, for Equiformer V2, our implementation of the analytical Hessian required approximately six times as many node hours as performing all the inference calls for the finite differences method.
This result should depend on the number of adsorbate atoms being displaced however, since the finite differences approach adds six displacement dimensions per atom.
This result is surprising, but after troubleshooting we were unable to determine an obvious culprit, therefore we elected to use numerical Hessians derived from finite differences for all \gls{gnn} potential experiments in this work.
If a more efficient approach to calculating the analytical Hessian on \gls{gnn} potentials could be implemented (particularly for Equiformer V2), with compute time on the order of a single inference call, then our results suggest this would be a favorable approach to computing the Hessian for \gls{gnn} potentials.
As it stands, the compute costs of performing inference on \gls{gnn} potentials is substantially lower than resolving \gls{vasp} calculations, therefore we are satisfied that the numerical approach is currently worth using.

\subsection{Validation of CPES with ML}
\begin{figure}
    \centering
    \includegraphics[width=0.95\linewidth]{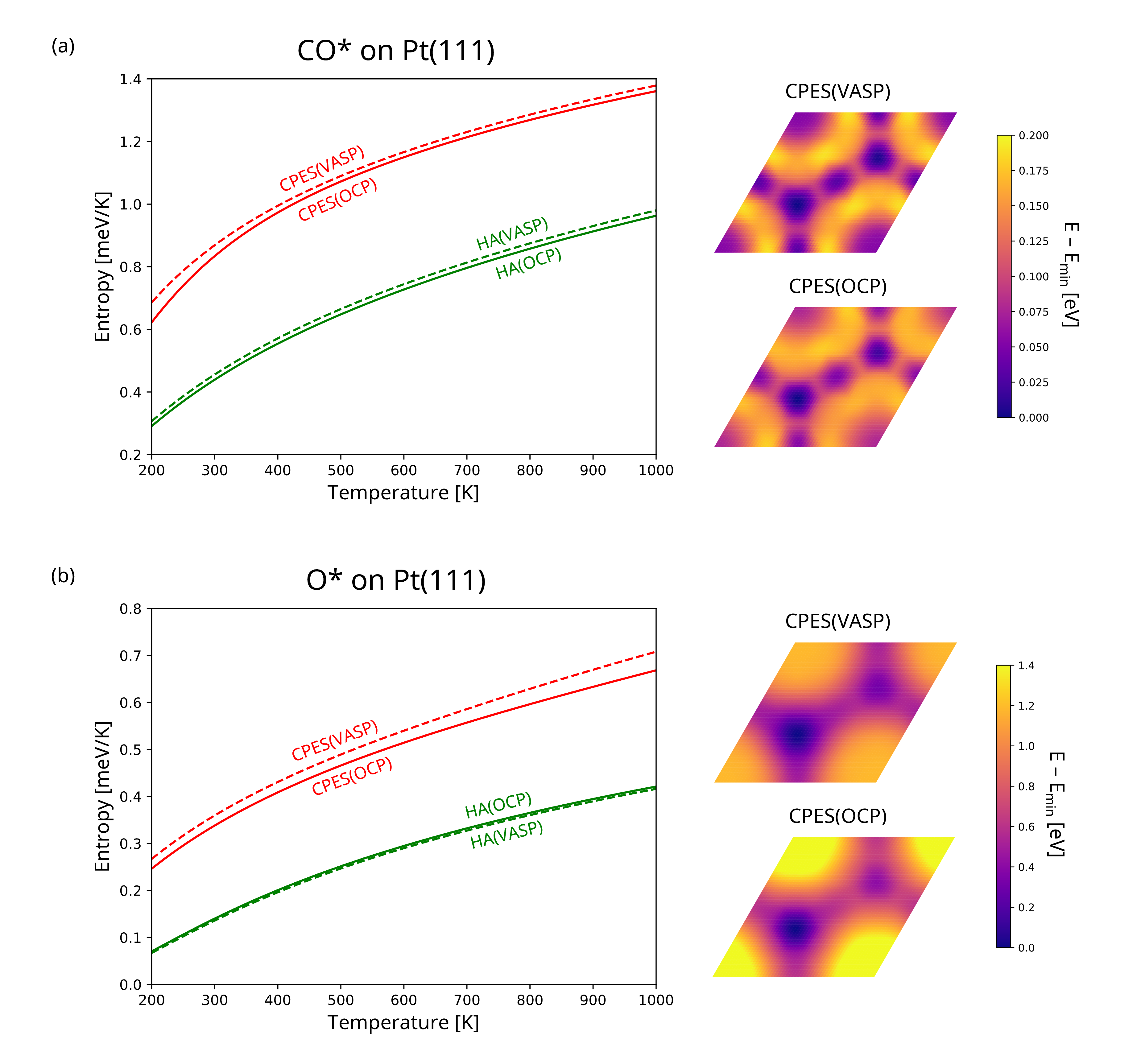}
    \caption{Plots of entropy vs temperature for CO* (a) and O* (b) on Pt(111), obtained with the harmonic approximation (HA) and the complete potential energy sampling (CPES) methods, and using energies calculated with VASP or the pretrained OCP ML potential. The 2D plots of potential energy surfaces are also shown in the figure.
    }
    \label{fig:cpes-validation}
\end{figure}

\end{document}